\documentstyle[11pt,epsfig]{article}
\begin{document}

\title{Electron spin dynamics in quantum dots and related nanostructures due
to hyperfine interaction with nuclei}
\author{John Schliemann, Alexander Khaetskii, and Daniel Loss\\\\
Department of Physics and Astronomy, University of Basel,\\
CH-4056 Basel, Switzerland}

\date{\today}

\maketitle

\abstract{We review and summarize recent theoretical and experimental work
on electron spin dynamics in quantum dots and related nanostructures due to
hyperfine interaction with surrounding nuclear spins. 
This topic is of particular interest with respect to several
proposals for quantum information processing in solid state systems.
Specifically, we investigate the hyperfine interaction of an electron spin
confined in a quantum dot in an s-type conduction band with
the nuclear spins in the dot. This interaction is proportional to the
square modulus of the electron wave function at the location of
each nucleus leading to an inhomogeneous coupling, i.e. 
nuclei in different locations are coupled with different strength.
In the case of an initially fully polarized nuclear spin system an
exact analytical solution for the spin dynamics can be found. For
not completely polarized nuclei, approximation-free results can only be 
obtained numerically in sufficiently small systems. We compare these
exact results with findings from several approximation strategies.}

\section{Introduction}

In the recent years an extraordinary and increasing interest in 
spin-dependent phenomena in semiconductors has
developed in the solid state physics community \cite{Wolf01,Awschalom02}. 
These research activities are
often labeled by the keyword ``spintronics'' which summarizes 
the entire multitude of efforts towards using the electron spin
rather than, or in combination with, its charge for information
processing, or, even more ambitious, quantum information processing.
In fact, in the recent years a series of proposals for implementing
quantum computation in solid state systems using electron and/or nuclear
spins have been put forward \cite{Awschalom02,Loss98,Privman98,Kane98,Barnes00,Levy01,Ladd01}.
In order to use the electron spin as an information carrier, long spin 
decoherence times are desirable if not indispensable.
A serious possible limitation of spin coherence in semiconductors is the
hyperfine interaction with surrounding nuclear spins. In fact, in
semiconductors isotopes carrying a nonzero magnetic moment are ubiquitous.
The commercial use of semiconductor technology is so far grossly dominated
by silicon applications. In this material the magnetic isotope $^{29}{\rm Si}$
having a spin 1/2 and a magnetic moment of $-0.5553$ nuclear magnetons
occurs with a natural abundance of $4.7\%$, apart from the two spinless
stable isotopes $^{28}{\rm Si}$ and $^{30}{\rm Si}$. From these numbers,
hyperfine interaction might not appear to be a particularly relevant
issue. However, the systems presently mostly studied in the field
of spinelectronics and solid-state quantum information processing contain
materials such as GaAs, (Ga,Al)As and InAs whose elements consist entirely of
spin-carrying isotopes with substantial magnetic moments. 
As it will be discussed in detail in the present review article, in such 
systems
the hyperfine coupling of electron spins to nuclear spins can easily become
an important interaction. In table \ref{table1} we summarize the natural
abundances and magnetic properties of stable spin-carrying nuclei relevant to
semiconductor systems.

A principle way to avoid such hyperfine couplings to electron spins is
to use isotopically purified material containing only a strongly
reduced amount of magnetic isotopes \cite{Haller95}. 
However, present technology allows
isotope purification of typical semiconductor materials only up
to a few hundredths of a percent or more of unwanted isotopes remaining.
This degree of purification might in general not be sufficient to meet the high
precision demands on implementations of quantum information processing.
Moreover, such isotopically purified materials have often quite high prices
such that they might not appear as a technologically viable option even
if the demands on precision are lower.

This review is organized as follows: In section \ref{dyn} we summarize
important experimental work on electron spin dynamics in semiconductor
nanostructures which have motivated the mostly theoretical studies
we are reviewing in this article. In the following section we describe
the basic physics of the hyperfine interaction between electron
and nuclear spins as it occurs in semiconductors. Section \ref{hyper} is
devoted to a detailed analysis of electron spin dynamics in quantum dots
due to hyperfine interaction with nuclear spins. After specifying the
details of the model in section \ref{modeling}, we discuss the ground state
and elementary properties of the underlying Hamiltonian in section
\ref{elementary}. There we also describe a full analytical solution for
the electron spin dynamics which can be obtained in the case of an initially 
fully polarized nuclear spin system. 
In section \ref{integ} we outline how to solve for the
eigenstates and energy levels of the system via the Bethe
ansatz technique relying on the integrability of the hyperfine coupling
Hamiltonian.
In section \ref{typestate} we introduce different types of
initial states for the nuclear spin system. These different types of 
initial states lead significantly different electron spin dynamics which
are described in section \ref{numdyn}. The results reported on there
are based on exact diagonalizations of the Hamiltonian for sufficiently
small systems. In section \ref{entanglement}
we discuss the intimate connection between the decay of the electron spin
and the generation of quantum entanglement between the electron spin
and the nuclear spin system. Section \ref{ensemble} is devoted to the
important question of spin dephasing in an emsemble of dots as
opposed to decoherence of of a single electron spin.
In section \ref{approx} we discuss
several further approximate treatments of the electron spin dynamics 
that have appeared
in the recent literature and compare them with the (quasi-)approximation-free
approaches described before. 
Further, mostly theoretical, work relevant to the issue
of hyperfine interaction between an electron spin bound to a quantum dot
and surrounding nuclear spins is summarized in section \ref{further}. 
We close with conclusions and an outlook in section \ref{concl}.

\section{Electron spin dynamics in quantum dots}
\label{dyn}

The dynamics of electron spins confined in semiconductor nanostructures
is a rich and very active field. Among the most notable developments are
experiments by Kikkawa and Awschalom who demonstrated very remarkably
large coherence times for electron spins in n-doped bulk
GaAs \cite{Kikkawa98,Kikkawa99}. This time scale can exceed $100{\rm ns}$
and represents the $T_{2}^{*}$ time. i.e. the transverse spin relaxation
time of an {\em ensemble} of electrons. These experimental findings
have generated a great deal of prospects in the fields of spin electronics
and semiconductor quantum computing \cite{Wolf01,Awschalom02}.

A photoluminescence study of excitons localized in single GaAs quantum dots 
was done by 
Gammon {\it et al.} \cite{Gammon96}. In a subsequent study of the 
magnetooptical spectra of individual localized excitons \cite{Gammon01} 
the role of electron spin hyperfine interaction with nuclear spins
was investigated. Though these experiments dealt with {\it single} 
GaAs quantum dots (not ensembles of them), it was not the 
electron spin dynamics studied there, but spectra of individual 
localized excitons where the strong Coulomb interaction 
between an electron and a heavy hole is important. 
The relaxation lifetime of electron spins for an ensemble
of CdSe quantum dots of very small diameter (20-80 $\AA$)
 was measured by Gupta {\it et al.} using a femtosecond-resolved Faraday
rotation technique \cite{Gupta99}. Again, due to the small size of the dot  
the Coulomb interaction between electron and hole is not neglegible. 
Therefore, it is highly probable that the observed short relaxation time 
(being of order of nanoseconds) is due to fast spin dynamics of the hole. 
The surface states could also contribute to the spin relaxation due to the 
small size of the dot.
Finally Epstein {\it et al.} have analyzed
the spin lifetime of photogenerated carriers in InAs quantum dots
using the Hanle effect \cite{Epstein01}.

The above experiments \cite{Gupta99,Epstein01} were performed on 
{\it ensembles} of quantum dots,
not single dots. Therefore the estimated spin relaxation times are
$T_{2}^{*}$ time scales. The experiments which directly probe the single 
electron spin relaxation were done recently 
by Fujisawa {\it et al.} \cite{Fujisawa02} and by Hanson 
{\it et al.} \cite{Hanson03}.
The non-equilibrium tunneling current through excited states in an 
AlGaAs/GaAs quantum dot was studied
using a pulse-excitation technique which measures the energy relaxation time 
from
the excited state to the ground state. Very low spin-flip rates were observed 
which is consistent with the theoretical predictions 
\cite{Burkard99,Khaetskii00,Khaetskii01}.
The physical mechanisms for electron spin relaxation for delocalized states 
include the interplay of spin-orbit coupling with impurity scattering 
and/or electron-phonon interaction, and the
hyperfine interaction with surrounding nuclear spins.
In recent theoretical studies Khaetskii and Nazarov have
concluded that the first type of mechanisms is strongly suppressed
for electrons localized in quantum dots \cite{Khaetskii00,Khaetskii01}, 
see also Ref.\cite{Halperin01}. 
As to the contribution to the electron spin {\it decoherence} due to the 
combined effect 
of the spin-orbit interaction and the spin-independent interaction with 
acoustic phonons,
there is an indication that within the Markovian approximation, which is 
usually applicable to this problem, $T_2$ time can be as long as $T_1$ time. 
This is  due to the fact that an additional  contribution to the $1/T_2$ 
rate originating from the fluctuations of the spin-orbit related magnetic 
field along  the external magnetic field direction 
is proportional to the phonon density of states at zero frequency. Thus, 
this contribution is zero for acoustic phonons, see Ref.\cite{Khaetskii03}.  
These results have motivated
a whole variety of theoretical investigations on hyperfine interaction
in quantum dots and related structures which will be reviewed in this article
\cite{Erlingsson01,Khaetskii02a,Khaetskii02b,Khaetskii03,Lyanda-Geller02,Erlingsson02,deSousa02a,deSousa02b,Merkulov02,Saykin02,Schliemann02,Eto02,Semenov03a,Saikin02,Pershin03a,Taylor03,Imamoglu03,Pershin03b,Semenov03b}.

A scenario similar to an electron bound in a quantum dot is the case
of a shallow phosphorus donor in a silicon crystal, Si:P. 
This case is essential for the solid state quantum computing proposal
by Kane \cite{Kane98}. The donor electron is bound to the P atom in a 
large hydrogen-like orbit with a Bohr radius of about $30{\rm \AA}$.
The nuclear spins interacting with the electron are the central
$^{31}{\rm P}$ and the surrounding $^{29}{\rm Si}$.
The $T_{1}$ time scale for energy relaxation of the electron spin
was determined by Feher and Gere to be of order $10^{3}{\rm s}$
\cite{Feher59}. The transverse spin decay was investigated by 
Gordon and Bowers \cite{Gordon58}
using the spin-echo method deducing a time scale
$T_{2}^{*}$ of order $500{\rm \mu s}$, see also \cite{Chiba72}. A very recent
spin-echo study by Tyryshkin {\it et al.} \cite{Tyryshkin03} on
P donors in natural and isotopically purified Si has reported 
$T_{2}^{*}$ times being significantly larger than the previous result.

Finally we mention that hyperfine interaction between electron 
and nuclear spins
is of course also investigated in higher-dimensional semiconductor
nanostructures such as quantum wells. For recent work using optical
NMR techniques we refer to \cite{Salis01,Eickhoff02}.

\section{Hyperfine interaction in semiconductors}
\label{Fermi}

Hyperfine interaction is the coupling of a nuclear magnetic moment to 
the magnetic field provided by the (orbital and spin) magnetic moment
of electrons. The Hamiltonian describing this interaction in the
as a lowest-order relativistic correction
was derived in 1930 by Fermi \cite{Fermi30}.
For an $s$-electron there is no orbital contribution, and the Hamiltonian reads
\cite{Fermi30,Pikus84}
\begin{equation}
{\cal H}=\frac{4\mu_{0}}{3I}\mu_{B}\mu_{I}|\psi(\vec r_{I})|^{2}\vec S\vec I\,.
\label{hyperham}
\end{equation}
Here $\vec S$ is the spin of the electron, and $\psi(\vec r_{I})$ is its wave 
function at the location $\vec r_{I}$
of the nucleus. This Hamiltonian couples the
electron spin to the nuclear spin $\vec I$ with total spin quantum number 
$I$ and magnetic moment $\mu_{I}$, which is represented by the
operator $\vec \mu_{I}=(\mu_{I}/I) \vec I$.
Both spin operators are taken to be dimensionless,
$\mu_{0}=4\pi\cdot 10^{-7}{\rm Vs}/{\rm Am}$ is the usual magnetic
constant in SI units, and $\mu_{B}$ is the Bohr magneton.
The leading contribution to the hyperfine coupling for
higher angular momenta of the electron looks different from (\ref{hyperham})
and is essentially given by the usual dipolar coupling between the
nuclear magnetic moment and the (orbital and spin) magnetic moment of the
electron.

For the issue of spin coherence and interaction with nuclei in semiconductors,
the most relevant case are electrons in $s$-type conduction bands.
We therefore shall concentrate on this case where the hyperfine coupling is
described by the Hamiltonian (\ref{hyperham}). From the viewpoint of
nuclear magnetic resonance experiments this coupling is the origin of an
increment in the position of resonance lines known as the Knight shift. From 
the viewpoint of electron spin resonance effects in solids this
Hamiltonian describes the Overhauser field. As a general reference on both
effects we refer the reader to the textbooks by Abragam \cite{Abragam61}
and by Slichter \cite{Slichter90}.

In a semiconductor crystal
the electron wave function is a product of a Bloch amplitude
$u(\vec r)$ and a modulating envelope function $\Psi(\vec r)$,
$\psi(\vec r)=\Psi(\vec r)u(\vec r)$. We therefore can rewrite the
Hamiltonian as \cite{Pikus84}
\begin{equation}
{\cal H}=\frac{4\mu_{0}}{3I}\mu_{B}\mu_{I}
\eta|\Psi(\vec r_{I})|^{2}\vec S\vec I
\label{hambloch}
\end{equation}
with $\eta=|u(\vec r_{I})|^{2}$. For free electrons the Bloch function is
constant, $|u(\vec r)|=1$; 
in a realistic crystal $|u(\vec r)|$ has maxima at the 
lattice positions, i.e. the locations of the nuclei, leading to $\eta>1$.
Values for $\eta$ in semiconductor systems can be estimated from
electron spin resonance experiments \cite{Pikus84}. For InSb Gueron 
\cite{Gueron64} found $\eta_{\rm In}=6.3\cdot 10^{3}$, 
$\eta_{\rm Sb}=10.9\cdot 10^{3}$;
for GaAs Paget {\it et al.} \cite{Paget77} estimated
$\eta_{\rm Ga}=2.7\cdot 10^{3}$, $\eta_{\rm As}=4.5\cdot 10^{3}$. From 
NMR experiments on $^{29}{\rm Si}$ Shulman and Wyluda estimated
a value of $\eta_{\rm Si}=186$ \cite{Shulman56}.

\section{Electron spin dynamics in quantum dots due to hyperfine interaction 
with nuclei}
\label{hyper}

\subsection{Modeling hyperfine interaction in quantum dots}
\label{modeling}

We consider an electron confined in a semiconductor quantum dot in an $s$-type
conduction band. We assume the electron to be in some orbital
eigenstate according to the confining potential, e.g. the orbital
ground state in the quantum dot. The remaining spin degree of freedom is
coupled to an external magnetic field $\vec B$ with an electronic
g-factor $g$, and to the spins of surrounding nuclei via the
the hyperfine contact interaction described in the previous section.
Thus the Hamiltonian reads
\begin{equation}
{\cal H}=g\mu_{B}\vec S\vec B+\vec S\sum_{i}A_{i}\vec I_{i}\,.
\label{defham}
\end{equation}
Here the subscript $i$ labels the nuclei, and the coupling constants
$A_{i}$ are given by (cf. (\ref{hambloch})) 
\begin{equation}
A_{i}=Av_{0}|\Psi(\vec r_{i})|^{2}
\label{singlehypercoupling}
\end{equation}
with 
\begin{equation}
A=\frac{4\mu_{0}}{3I}\mu_{B}\mu_{I}\eta n_{0}
\label{hypercoupling}
\end{equation}
where $n_{0}=1/v_{0}$ is the density of nuclei. Provided that the
electronic envelope wave function $\Psi(\vec r)$ varies smoothly on
the length scale given by $^{3}\sqrt{v_{0}}$ it is appropriate to
replace the sum $\sum_{i}A_{i}$ by an integral over space; 
then $A=\sum_{i}A_{i}$ up to small corrections to this approximation.
The Hamiltonian (\ref{defham}) also describes the hyperfine interaction
between nuclear spins and the spin of an electron bound in a hydrogen-type
orbit around a phosphorus donor in a silicon crystal. 

Let us now address the order of magnitude of the hyperfine interaction.
A GaAs quantum dot with a volume of order $10^{5} {\rm nm}^{3}$ contains 
of order $N=10^{6}$ nuclei with a density $n_{0}=45.6 {\rm nm}^{-3}$.
Taking into account the natural abundances of the three occurring isotopes
$^{69}{\rm Ga}$, $^{71}{\rm Ga}$, $^{75}{\rm As}$ one has an average
nuclear magnetic moment of $\mu_{I}=1.84 \mu_{N}$. With the values for
$\eta$ estimated in \cite{Paget77} this leads to an
overall coupling constant $A$ of order $10^{-5}{\rm eV}$ to $10^{-4}{\rm eV}$.
This is the
strength of the hyperfine coupling acting on the electron spin
in the presence of a fully polarized nuclear spin system. With the
effective electron g-factor $g=-0.44$ for GaAs this corresponds to an effective
magnetic field of order a few Tesla. For a completely unpolarized nuclear
spin system the strength of the hyperfine field is fluctuating around zero
with a typical value given by $A/\sqrt{N}$ where $N$ is the number of
nuclei in the dot effectively interacting with the electron spin. Note that
for GaAs all nuclear magnetic moments are positive, leading to an
antiferromagnetic sign for the hyperfine coupling. This is different 
from the situation in Si:P where the magnetic moment of $^{31}{\rm P}$
is positive while the surrounding $^{29}{\rm Si}$ have a negative nuclear
magnetic moment, resulting in a dominantly ferromagnetic coupling to 
the electron spin. The natural abundance of $^{29}{\rm Si}$ 
leads to a density $n_{0}=2.3{\rm nm}^{-3}$. Assuming a Bohr radius
of $30 {\rm \AA}$ for the hydrogen-like electron orbit one can estimate the 
number of nuclei effectively interacting with the electron spin to be of 
order a few hundred, and from (\ref{hypercoupling}) one finds
values of $|A|$ of order  $10^{-7}{\rm eV}$ using the estimate for
$\eta$ given in \cite{Shulman56}. 

Another kind of interaction in the systems discussed above is the dipolar
coupling between nuclear spins. This contribution to the Hamiltonian
is obtained from its classical counterpart by expressing the
magnetic moments in terms of nuclear spin operators,
$\vec \mu_{I}=(\mu_{I}/I)\vec I$. Thus the interaction between two nuclear 
spins labeled by 1 and 2 reads
\begin{equation}
{\cal H}_{12}=-\frac{\mu_{0}}{4\pi}
\frac{\mu_{1}\mu_{2}}{I_{1}I_{2}}\frac{1}{r_{12}^{3}}
\left(
\frac{3(\vec I_{1}\vec r_{12})(\vec I_{2}\vec r_{12})}{r_{12}^{2}}
-\vec I_{1}\vec I_{2}\right)
\label{dipolar}
\end{equation}
where $\vec r_{12}$ is the distance vector between the nuclei. In GaAs
$^{71}{\rm Ga}$ has the largest nuclear magnetic moment with
$\mu_{I}=2.562\mu_{N}$. Assuming two of such nuclei being on
nearest neighbour sites of the underlying zinc-blende lattice one finds
$r_{12}=0.24{\rm nm}$, and the energy scale of the dipolar interaction
between these two nuclei has the value 
$(\mu_{0}/4\pi)\mu_{I}^{2}/r_{12}^{3}=7.6\cdot 10^{-12}{\rm eV}$.
This is an upper bound for the dipolar coupling 
between neighboring nuclei in GaAs and 
sets the time scale ($10^{-4}\div 10^{-5} {\rm s}$) on 
which this interaction indirectly 
influences the quantum dynamics of the electron spin.
As we shall see below, the largest time scales relevant for electron
spin dynamics due hyperfine coupling are of order $10^{-6}{\rm s}$.
Therefore, the time scale of the dipolar coupling is much larger than the 
time scale provided by the hyperfine interaction; analogous 
considerations can be made for the case of Si:P.
In the following we shall therefore neglect
the dipolar interaction unless stated otherwise. A contribution to the 
Hamiltonian which we also neglect is the coupling of the nuclear spins 
to the external magnetic field. This interaction is much smaller than the
Zeeman coupling of the electron spin because of the smallness of the nuclear
magneton compared with the Bohr magneton for electrons.

\subsubsection{Ground state and elementary properties}
\label{elementary}

\begin{figure}
\centerline{\includegraphics[width=8cm]{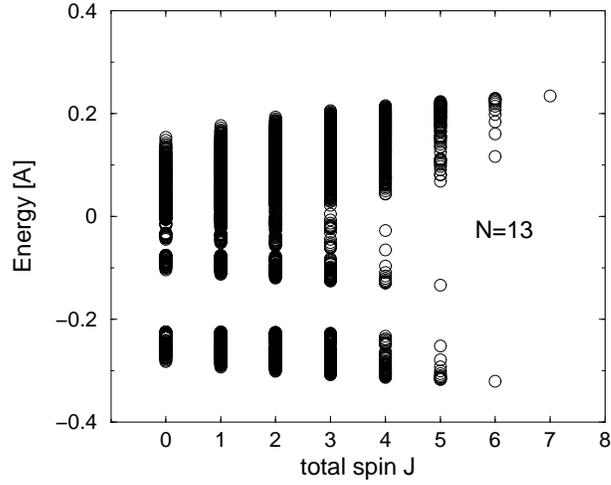}}
\caption{Energy spectrum (in units of $A>0$) of the Hamiltonian
(\ref{defham}) for zero magnetic field as a function of the total spin 
quantum number $J$ in a system of $N=13$ nuclear spins of length $1/2$. 
For details about the modeling of coupling constants see section 
\ref{technical}.
\label{spectrum}}
\end{figure}

If the sign of the hyperfine interaction is ferromagnetic ($A<0$, 
as realized by the coupling to $^{29}{\rm Si}$ nuclear spins)
the ground state of the Hamiltonian (\ref{defham}) is 
(for $\vec B=0$) the fully spin-polarized
multiplet with all (electron and nuclear) spins in parallel resulting in
the maximum value of the total spin quantum number, $J=NI+1/2$.
For an antiferromagnetic sign the classical ground state
(neglecting the operator nature of spins) has all nuclear spins in parallel
and the electron spin pointing opposite to them. We therefore anticipate
that the true quantum mechanical ground state will have 
(again for vanishing external magnetic field) the total spin 
quantum number $J=NI-1/2$. This assumption is confirmed by our numerics.
Fig.~\ref{spectrum} shows the energy spectrum of the Hamiltonian \ref{defham}
for zero magnetic field as a function of the total spin 
quantum number $J$ in a system of $N=13$ nuclear spins of length $I=1/2$. 
Clearly the ground state lies in the subspace of $J_z
=(13-1)/2=6$. 
A general state in this subspace can be written as
\begin{equation}
|\psi\rangle=\alpha S^{-}|\Uparrow,\uparrow\cdots\uparrow\rangle
+\sum_{i}\frac{\beta_{i}}{\sqrt{2I}}
I_{i}^{-}|\Uparrow,\uparrow\cdots\uparrow\rangle
\label{defstate}
\end{equation}
where $i$ labels the nuclei, $|\Uparrow,\uparrow\cdots\uparrow\rangle$
is the fully spin-polarized ground state with all spins pointing upward,
and we have introduced the usual spin lowering operators
for the electron spin of length $1/2$ and the nuclear spins
of length $I$. The
stationary Schr\"odinger equation leads to the following
system of equation for the amplitudes $\alpha$ and $\beta_{i}$:
\begin{eqnarray}
-\left(\frac{I}{2}A+\frac{\varepsilon_{z}}{2}+E\right)\alpha
+\sum_{i}\sqrt{\frac{I}{2}}A_{i}\beta_{i} 
& = & 0\\
\sqrt{\frac{I}{2}}A_{i}\alpha
+\left(\frac{I}{2}\left(A-2A_{i}\right)+\frac{\varepsilon_{z}}{2}-E\right)
\beta_{i}
& = & 0
\end{eqnarray}
Here $E$ is the energy eigenvalue, and we have reintroduced a finite
Zeeman coupling $\varepsilon_{z}=g\mu_{B}B$. From the second of the 
above equations one finds
\begin{equation}
\beta_{i}(E)=
\frac{\sqrt{I/2}A_{i}\alpha(E)}{E-(I(A-2A_{i})+\varepsilon_{z})/2}\,.
\label{beta}
\end{equation}
The ground state energy for $|\varepsilon_{z}|\ll A$ is of order 
$E\approx-AI/2$, thus the denominator of the r.h.s. of (\ref{beta})
is of order $A$. Since the couplings $A_{i}$ are of order $A/N$
we see that each $\beta_{i}$ is smaller than $\alpha$ by a factor 
of order $1/N$. Therefore $|\alpha|\approx 1$ up to quantum corrections, 
and all coefficients $\beta_{i}(E)$ are of order $1/N$. Thus, the 
corrections to the classical ground state ($|\alpha|=1$)
are of order $1/N$,
$|\alpha|^{2}=1-\sum_{i}|\beta_{i}|^{2}\approx 1-1/N$. 
In summary, for large systems ($N\gg 1$), the ground state is essentially
given by a tensor product state with all nuclear spins in parallel and the
electron spin pointing opposite to them. 
Excited  states with the same total spin quantum number along the direction 
of the nuclear polarization
are separated from the ground state by a substantial gap of the order of the 
coupling parameter $A$. 

\begin{figure}
\begin{center}
\leavevmode
\includegraphics[width=13cm]{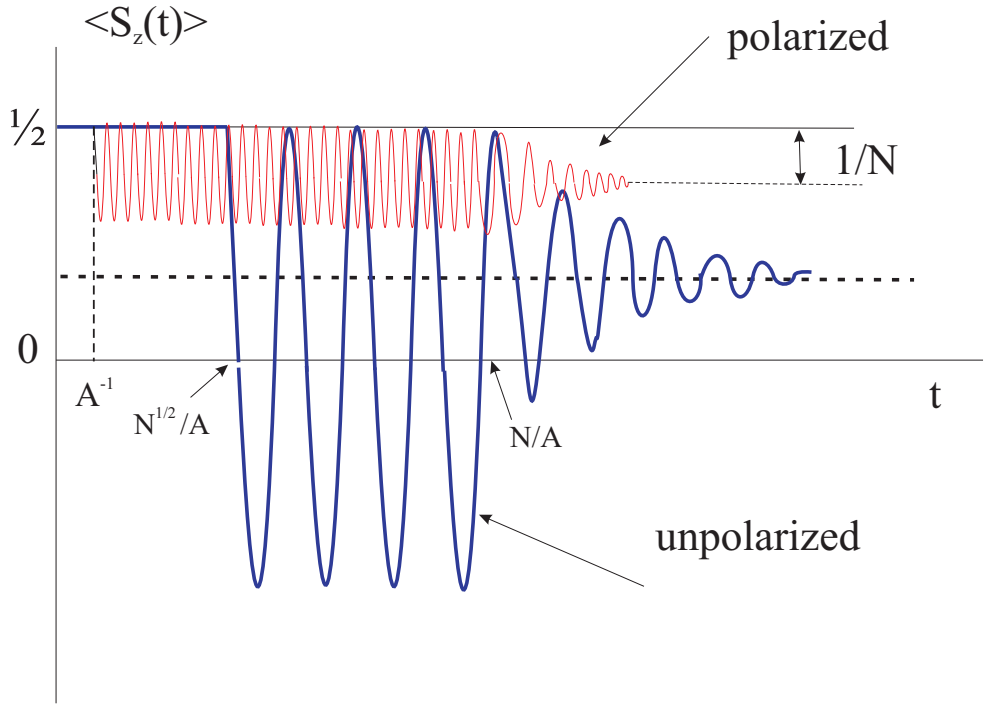}
\end{center}
\caption{Schematic  dependence of $<S_z(t)>$ on time t
for the unpolarized tensor product and fully polarized nuclear states. 
The time scale for the onset of the decay $\sim N/A$ is the same for both 
cases. In the fully polarized case the magnitude of the effect is $1/N$. 
The period of oscillations is of the order of $\sqrt{N}/A$ for the 
unpolarized and $\sim 1/A$ for the polarized case.}
\label{exsol1}
\end{figure}

These results were obtained recently by Khaetskii, Loss, and Glazman
\cite{Khaetskii02a} who studied the time evolution of the classical
ground state under the quantum Hamiltonian (\ref{defham}).
These investigations were carried out using the standard 
Laplace transform technique.  A  detailed account
on the mathematical details of this approach has been
given recently in \cite{Khaetskii02b}. 
The main findings are the following. The classical 
ground state for $|\varepsilon_{z}|\ll A$ remains constant in time 
up to quantum corrections of order $1/N$. Starting the time
evolution at $t=0$ with $|\alpha(t=0)|=1$ the electron spin expressed
in terms of its expectation value $\langle S^{z}(t)\rangle$
undergoes {\em coherent oscillations} between 
$\langle S^{z}(t)\rangle=-1/2$ and 
$\langle S^{z}(t)\rangle=-1/2+{\cal O}(1/N)$ with a period of
$T=4\pi\hbar/A$ over a time scale
of order $\hbar N/A$. This time scale is nothing but the characteristic 
period of precession of an individual nuclear spin in the field 
generated by the electron spin.  
At this time scale a different
nuclear spin configuration is  created, and because of the spatial variation 
of the hyperfine
coupling constants inside the dot, this
leads  to a different random value of the  nuclear  field seen by the
electron spin and thus to its decoherence.
After a time interval of this order 
the oscillations fade out and the expectation value 
$\langle S^{z}(t)\rangle$ decays to a value of order
$\langle S^{z}(t\to\infty)\rangle=-1/2+{\cal O}(1/N)$, see Fig.~\ref{exsol1}.
  
Thus the decaying part of the initial
spin state has smallness $1/N$ which is due to a large gap $\simeq A$ seen by 
the electron spin through the hyperfine interaction for a fully polarized 
state. As a result, only a small portion $\sim 1/N$ of the opposite (+1/2) 
state can be admixed. 
Moreover, the decay of the electron spin turns out
to be {\em nonexponential} for any external magnetic field. In the case 
when the perturbative treatment is applicable (i.e. a large Zeeman field)  
it follows a power law  with a leading term proportional to $t^{-3/2}$. 
The fact that the decay of the electron spin 
is not exponential can be easily understood. The exponential decay occurs 
when the correlation time of the randomly fluctuating field which causes 
the decoherence is short compared to the decoherence time, 
as a result  the Markovian approximation can be applied.  
In our case the  nonexponential behavior is a
result of the fact that the correlation time for the nuclear magnetic field 
seen by the electron spin is itself determined by the flip-flop processes 
since the internal nuclear dynamics is excluded. Thus the Markovian 
approximation is not valid, see for comparison section \ref{Markovian}.   

\begin{figure}
\begin{center}
\leavevmode
\includegraphics[width=13cm]{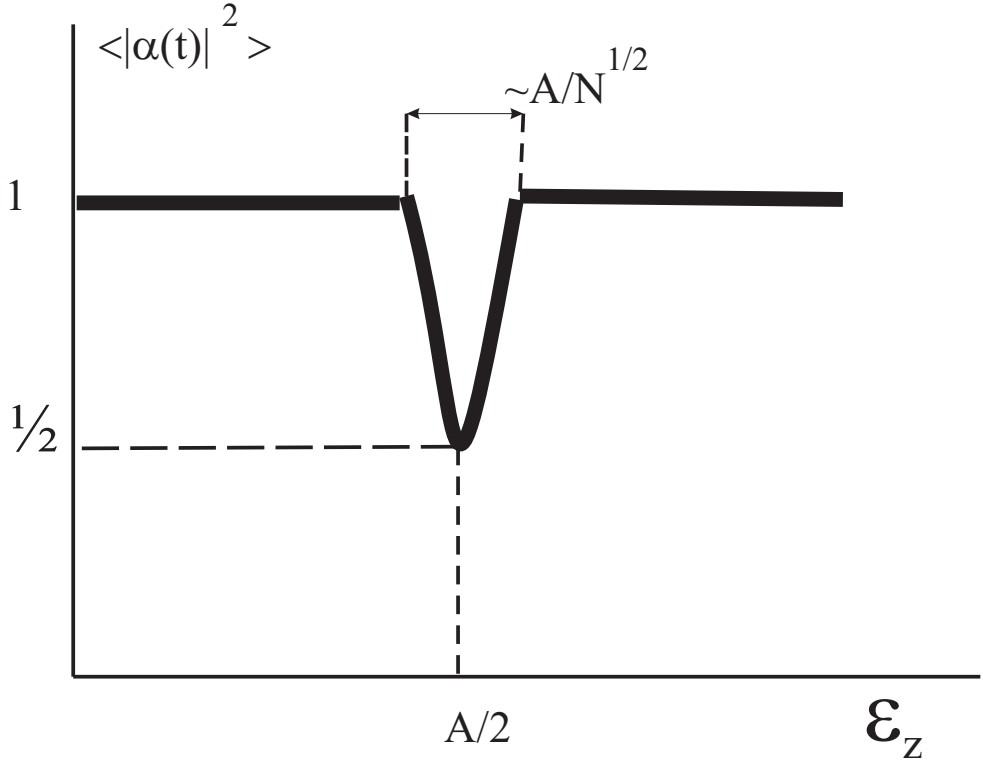}
\end{center}
\caption{The dependence of $|\alpha|^2$ averaged over time 
($<|\alpha(t)|^2>$) on the external Zeeman field $\epsilon_z$ for a 
fully polarized nuclear state. The resonance occurs at $|\epsilon_z| = A/2$, 
and the width of the resonance is $\sim A/\sqrt{N}$, which is much smaller 
than the initial gap $A/2$.}
\label{exsol2}
\end{figure}

A particular situation arises in the time evolution of the state
(\ref{defstate}) with initially $|\alpha(t=0)|=1$ if an external Zeeman
field is applied to the electron spin which approximately cancels the initial
Overhauser field, i.e. $\varepsilon_z \approx -AI$ in the above conventions
\cite{Khaetskii02a}.  Near this Zeeman field 
$|\alpha|^2$ averaged over time  is 1/2, i.e. the
up and down states of the electron spin are strongly coupled via the nuclei 
(see Fig.~\ref{exsol2}). In contrast,
outside this resonance regime  the value  of $|\alpha|^2$ is
close to unity (with small $1/N$ corrections), i.e. 
$\langle S^z(t)\rangle =1/2-|\alpha|^2$ is close to
-1/2 at any time. The width of the resonance is $\sim A/\sqrt{N}$, i.e. small 
compared to
the initial gap $AI$.   This abrupt change  in the 
amplitude of oscillations of $\langle S^z(t)\rangle$ 
(when changing $\varepsilon_z$ in
a narrow interval around $AI$) can be used for an experimental detection of 
the fully polarized state. 

Although physically not particularly realistic, it is also instructive
to study the model (\ref{defham}) in the case of all coupling constants
being equal to each other, $A_{i}=A/N$ for all $i$. Models of this
type were studied recently by Eto \cite{Eto02} and by Semenov and Kim
\cite{Semenov03a}, see also \cite{Khaetskii02b}. 
The technical advantage of this type of
models lies in the fact that the square of the total nuclear spin
$\vec I_{tot}=\sum_{i}\vec I_{i}$ is an additional conserved quantity,
$[{\cal H}, I^{2}_{tot}]=0$. 
 If $A_{i}=A/N$ 
the Hamiltonian reads in the absence of an external
magnetic field
\begin{equation}
{\cal H}=\frac{A}{2N}\left(\left(\vec I_{tot}+\vec S\right)^{2}
-\vec I^{2}_{tot}-\vec S^{2}\right)\,.
\end{equation}
Since the total spin $\vec J=\vec I_{tot}+\vec S$ can have values
$J=I_{tot}\pm 1/2$, each value of the quantum number $I_{tot}$ corresponds to
two energy levels given by 
\begin{equation}
E(I)=\frac{A}{2N}
\left(\pm\left(I_{tot}+\frac{1}{2}\right)-\frac{1}{2}\right)\,.
\label{ergI}
\end{equation}
These energy levels are typically highly degenerate.
For instance, if the nuclear spins are of length $1/2$,
the different values of $I_{tot}$ occur with a degeneracy of
\begin{equation}
{N\choose{N/2-I_{tot}}}-{N\choose{N/2-I_{tot}-1}}\,.
\end{equation}
As seen from Eq.~(\ref{ergI}), the spectrum is equidistant with a level spacing
$\Delta E=A/2N$. Therefore the time evolution of an arbitrary state
is strictly periodic with a recurrence time $T=4\pi\hbar N/A$. However,
in contrast to naive expectations even in this simple case there is some time 
dependence of $\langle S^z(t)\rangle$ which cannot be described by a single 
frequency. Actually for a  nuclear state with given $I^z_{tot}$
the solution contains  all the frequencies of the form 
$\Delta E=A(2I_{tot}+1)/2N$, where $I_{tot}$ are all the moduli of the 
total nuclear momentum which can have this projection $I^z_{tot}$. 
  
Depending on the number and length of the nuclear spins, this can
lead to shorter periodicities in the time evolution of 
$\langle\vec S(t)\rangle$. For instance, for an odd number of half-integer
nuclear spins $2I_{tot}+1$ is even, and $\langle\vec S(t)\rangle$ has a
period of $T=2\pi\hbar N/A$

\subsubsection{Integrability}
\label{integ}

If all nuclear spins are of length $1/2$, the spin Hamiltonian (\ref{defham}) 
has the strong mathematical 
property of being integrable and solvable via an appropriate version of
the Bethe ansatz. This fact was recognized first by Gaudin in a rather
formal context \cite{Gaudin76}. Consider a system of $N$ spins $1/2$
and the following sequence of operators \cite{Gaudin76,Garajeu02}
\begin{equation}
{\cal H}_{i}=\sum_{j\neq i}\frac{\vec\sigma_{i}\vec\sigma_{j}}{z_{i}-z_{j}}
\end{equation}
where $\vec\sigma_{i}$ are Pauli matrices and the $z_{i}$ are some
arbitrary coupling parameters. Obviously, by fixing one spin $i$ to be the
central spin and adjusting the couplings to the other spins, the
operator ${\cal H}_{i}$ assumes the form of the Hamiltonian (\ref{defham}).
The operators ${\cal H}_{i}$ commute with each other,
\begin{equation}
\left[{\cal H}_{i},{\cal H}_{j}\right]=0\,,
\end{equation}
and fulfill the sum rule
\begin{equation}
\sum_{i}{\cal H}_{i}=0\,.
\end{equation}
Thus, $N-1$ of these operators together with the square of the total spin
form a set of $N$ linearly independent commuting operators being
bilinear in the individual spin operators. The coordinate Bethe ansatz 
diagonalizing simultaneously all ${\cal H}_{i}$ can be summarized
as follows \cite{Gaudin76,Garajeu02}. Consider states of the form
\begin{equation}
|w_{1},\dots,w_{m}\rangle=F(w_{1})\cdots F(w_{m})|\uparrow\cdots\uparrow\rangle
\label{bethestate}
\end{equation}
where $|\uparrow\cdots\uparrow\rangle$ is the spin-polarized state
with all spins in parallel, and
\begin{equation}
F(w)=\sum_{i}\frac{\sigma_{i}^{-}}{w-z_{i}}
\end{equation}
with $\sigma_{i}^{-}=\sigma_{i}^{x}-i\sigma_{i}^{y}$. These states are
eigenstates of all ${\cal H}_{i}$ if and only if the complex parameters
$w_{1}\dots w_{m}$ fulfill the {\em Bethe equations}
\begin{equation}
\sum_{i=1}^{N}\frac{1}{w_{k}-z_{i}}
+\sum_{{l=1}\atop{l\neq k}}^{m}\frac{2}{w_{l}-w_{k}}=0\,.
\label{betheeq}
\end{equation}
For a given solution of these equations the corresponding eigenvalues read
\begin{equation}
\varepsilon_{i}(w_{1},\dots,w_{m})=\sum_{l=1}^{m}\frac{2}{w-z_{i}}
+\sum_{j\neq i}\frac{1}{z_{i}-z_{j}}\,.
\end{equation}
The eigenstates (\ref{bethestate}) are states of highest weight in
multiplets of the total spin with respect to its $z$-component.
States lower in these multiplets can be obtained by applying the
lowering operator of the total spin, or by formally considering
solutions to the Bethe equations (\ref{betheeq}) with some $w_{k}$ being
infinite. By counting the number of solutions of
(\ref{betheeq}) it can be shown that the Bethe absatz produces all
multiplets, i.e. all energy eigenvalues.

The coordinate Bethe ansatz outlined above has been extended to the technique
of the algebraic Bethe ansatz \cite{Sklyanin96}. We also note that the
equations (\ref{betheeq}) are a limiting case of the Bethe equations
of the so-called Richardson model describing electron pairing in 
superconducting grains \cite{Richardson64}. This issue has attracted 
considerable interest recently 
\cite{Amico01,Schechter01,vonDelft02,Yuzbashyan03}, and an 
algebraic version of the Bethe ansatz has also been presented
\cite{vonDelft02}.

For practical purposes, the treatment of the Bethe equations
(\ref{betheeq}) is still rather complicated, and explicit results,
for instance for correlation functions, are difficult to obtain
\cite{Sklyanin99}. In the remainder of this review we shall therefore
concentrate on techniques other than the Bethe ansatz.
It is an interesting and important question to what extend
certain results such as the non-exponential spin decay
depend on the integrability of the model
(\ref{defham}). We stress that this integrability holds very generally and 
does not depend on a specific choice of the hyperfine coupling
constants $A_{i}$. 

\subsection{Different types of initial states}
\label{typestate}

In the numerical simulations to be described below the electron spin
is initially in a single tensor product state with the nuclear spin system,
\begin{equation}
|\psi(t=0)\rangle=|\psi_{el}\rangle\otimes|\psi_{nuc}\rangle\,,
\end{equation}
i.e. the electron spin described by $|\psi_{el}\rangle$ is initially 
uncorrelated with the nuclear spins. However, there is still quite a variety
of possibilities for the initial nuclear spin state $|\psi_{nuc}\rangle$.
A simple choice would be just a tensor product of eigenstates with
respect to a given quantization axis, say, the $z$-direction,
\begin{equation}
|\psi_{nuc}\rangle=|\uparrow\rangle_{1}\otimes|\downarrow\rangle_{2}
\otimes|\downarrow\rangle_{3}\otimes\cdots\otimes|\uparrow\rangle_{N}\,,
\label{updowntensor}
\end{equation}
where we have for simplicity assumed the nuclear spins to be of length $1/2$.
A nuclear spin state close to the above form can be generated experimentally
by cooling down the nuclear spins in a strong external magnetic field.
The strong magnetic field provides a quantization axis and suppresses
dipolar interactions changing the spin projection along the field axix.
Then, due to spin-lattice relaxation processes, the nuclear spin system will
end up in a state of the type (\ref{updowntensor}).

A more general possibility of a nuclear spin state is
a tensor product state but with
individual polarization direction for each spin,
\begin{equation}
|\psi_{nuc}\rangle=
\left(a_{1}|\uparrow\rangle_{1}+b_{1}|\downarrow\rangle_{1}\right)
\otimes\cdots\otimes
\left(a_{N}|\uparrow\rangle_{N}+b_{N}|\downarrow\rangle_{N}\right)\,,
\label{generaltensor}
\end{equation}
where the complex numbers $a_{i}$, $b_{i}$ parametrize the spin state
and are subject to the normalization condition $|a_{i}|^{2}+|b_{i}|^{2}=1$.
A yet more general state of the nuclear spins is a superposition of
tensor product states, 
\begin{equation}
|\psi_{nuc}\rangle=\sum_{T}\alpha_{T}|T\rangle\,,
\label{corr}
\end{equation}
where the summation runs over all tensor product states of the form
(\ref{updowntensor}), i.e. over a complete basis of the underlying Hilbert 
space. If one works in a subspace corresponding to a given value
of $J^{z}$, this summation has to be accordingly restricted.
For a nontrivial choice of the amplitudes $\alpha_{T}$ states of the
above form can in general not be expressed as tensor
product states, whatever basis one would choose in the space of each nuclear
spin. Therefore, such states are in general {\em correlated}, or,
using the language of quantum information theory, {\em entangled}
\cite{Schrodinger35,Nielsen00}.
A particular class of correlated nuclear states is obtained when the
complex amplitudes $\alpha_{T}$ are {\em chosen at random}, only
restricted by the normalization condition $\sum_{T}|\alpha_{T}|^{2}=1$.
In the following we will refer to this type of states as {\em randomly
correlated states}. As we shall see, the type of initial state of the
nuclear spin system has a profound impact on the electron spin dynamics.

\subsection{Numerical results for electron spin dynamics:
product states versus randomly correlated states}
\label{numdyn}

Here we review our recent numerical studies \cite{Schliemann02} 
of electron spin dynamics
in quantum dots modeled by the Hamiltonian (\ref{defham}).
These investigations are based on exact numerical diagonalizations 
whose technical details we summarize below.

\subsubsection{Numerical method and modeling}
\label{technical}

Our simulations of the electron spin dynamics are performed by
numerically diagonalizing the Hamiltonian (\ref{defham}). The numerical
diagonalization makes use of the fact that the projection of the
total spin $\vec J=\vec S+\sum_{i}\vec I_{i}$ on the direction of the
external field (usually chosen as the $z$-axis) is a conserved quantity 
leading to a block-diagonal structure of the Hamiltonian matrix.
Therefore it is convenient to work in subspaces of a given value of
$J^{z}$. The Hamiltonian is then diagonalized within such a subspace,
and the time evolution of a given initial state is obtained from the
eigensystem data. For a numerically exact simulation of the time evolution
of a general initial state one generally needs the full eigensystem, i.e.
all eigenvalues and corresponding eigenvectors. If the initial state 
involves more than one of the above invariant subspaces of the Hamiltonian
the time evolution in the different subspaces can be superimposed.
We note that this method for solving for the quantum mechanical
time evolution is non-iterative and can therefore be extended to very large
times. On the other hand it requires the full eigensystem in 
a given invariant subspace of the Hamiltonian, and it is the dimensions
of these subspaces that limit our numerical investigations.
To reduce the numerical demands we will consider in the following 
nuclear spin of length $I=\frac{1}{2}$. 
The dimensions of the invariant subspaces increase with decreasing $J^{z}$ 
starting from the maximum value $J^{z}=(N+1)/2$ according to a binomial
distribution where $N$ is the number of nuclei considered.
The dimensions of the invariant subspaces become largest for the 
minimum value of $|J^{z}|\in\{0,1/2\}$.
For the latter case we could simulate the full time evolution for systems
with up to fourteen nuclear spins. 

In our simulations we will assume a spherical quantum dot using the following
specific modeling. A given number $N$ of nuclear spins
is contained in a sphere of radius 
\begin{equation}
R=\left(\frac{3N}{4\pi n_{0}}\right)^{\frac{1}{3}}
\end{equation}
where $n_{0}=1/v_{0}$ is the density of nuclei. The electron wave function is 
given by
\begin{equation}
|\Psi(\vec r)|^{2}=\left(\frac{1}{\pi (R/a)^{2}}\right)^{\frac{3}{2}}
e^{-\frac{r^{2}}{(R/a)^{2}}}
\end{equation}
where the parameter $a$ describes the confinement of the electron in the dot
due to an essentially harmonic potential (with possibly small
anharmonic corrections).
In the following we shall use $a=2$ such that the electron is reasonably
confined in the sphere of radius $R$. The question of different types
of confining potentials was investigated in Ref.~\cite{Merkulov02}.
Moreover, we shall use the material parameters of GaAs with 
$n_{0}=45.6{\rm nm}^{-3}$. Therefore a typical quantum dot contains about
$N=10^{5}$ nuclei. To mimic their spherical distribution also in systems
of smaller size used in our simulations
we choose the radial coordinate $r_{i}$ of the $i$-th nucleus according to
\begin{equation}
r_{i}=\left(\frac{3(i-\frac{1}{2})}{4\pi n_{0}}\right)^{\frac{1}{3}}
\end{equation}
with $i$ ranging from $1$ to $N$. The results to be presented
below are obtained for an antiferromagnetic sign of the
hyperfine coupling, $A>0$. 

\subsubsection{Results for electron spin dynamics} 

\begin{figure}
\centerline{\includegraphics[width=8cm]{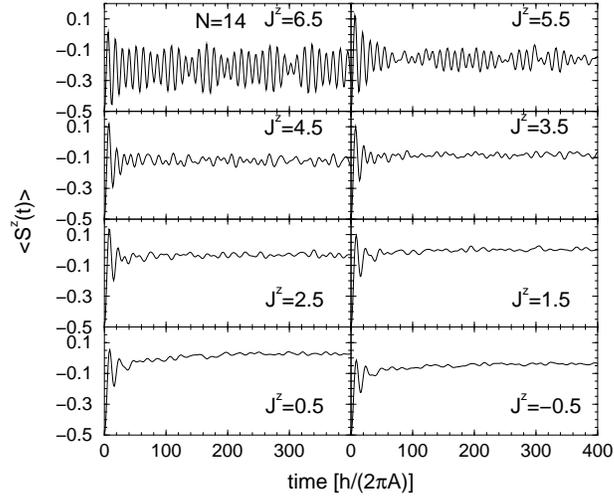}} 
\caption{The time evolution of the electron spin in a system of $N=14$ nuclear
spins of length $1/2$
for different degrees of polarization of the randomly
correlated nuclear system. The hyperfine coupling constants  are induced by the
quantum dot geometry. In all simulations the electron spin is
initially pointing downward in a tensor product with the nuclear system.
In the top left panel the nuclear spins are
fully polarized in the initial state with the electron spin
pointing opposite to them ($J^{z}=13/2$). In the following panels
the number of flipped nuclear spins in the initial state is
gradually increased. The case of an initially fully unpolarized
(but randomly correlated) nuclear system is reached in the bottom
right panel ($J^{z}=-1/2$). Here and in the following we take
spins to be dimensionless, i.e. measured in units of $\hbar$.
\label{dynamics1}}
\end{figure}

Fig.~\ref{dynamics1} shows results for a system of $N=14$ nuclear spins.
This is the largest system size for which we have been able to
treat the electron spin dynamics in the presence of an unpolarized
nuclear system. The upper left panel shows the expectation
value $\langle S^{z}(t)\rangle$ as a function of time for an initially
fully polarized nuclear system with the electron spin pointing opposite to
it in the negative $z$-direction. In the following panels 
the polarization of the nuclear system is successively reduced by
lowering the value of $J^{z}$ in the initial state. 
The case of a fully unpolarized nuclear spin system is reached in the bottom 
right panel with $J^{z}=-1/2$. Since the value of the $z$-component of the
total spin $J$ is fixed the expectation values of the transversal
components $\langle S^{x}\rangle$ and $\langle S^{y}\rangle$ vanish.
In all simulations shown in Fig.~\ref{dynamics1} the electron spin
is initially in a tensor product state with the nuclear system. The 
nuclear spins themselves are initially in a randomly correlated
state as described in section \ref{typestate}. 

In all cases,
$|\langle\vec S(t)\rangle|=|\langle S^{z}(t)\rangle|$
decreases in magnitude. With decreasing
polarization the decay becomes more pronounced, and the
oscillations accompanying this process get suppressed. Note that
it is the decay of the envelope in these graphs but not the 
fast oscillations themselves that signals the decay of the spin.
The distance between two neighboring maxima of the oscillations
can depend slightly on the initial state and the coupling constants
in the Hamiltonian. However, a good estimate for this effective period is
usually given by $T=4\pi\hbar/A$ since $A/2$ is an estimate
(neglecting quantum fluctuations) for the width of the spectrum,
i.e. the difference between the largest and the smallest eigenvalue 
of the Hamiltonian.

\begin{figure}
\centerline{\includegraphics[width=8cm]{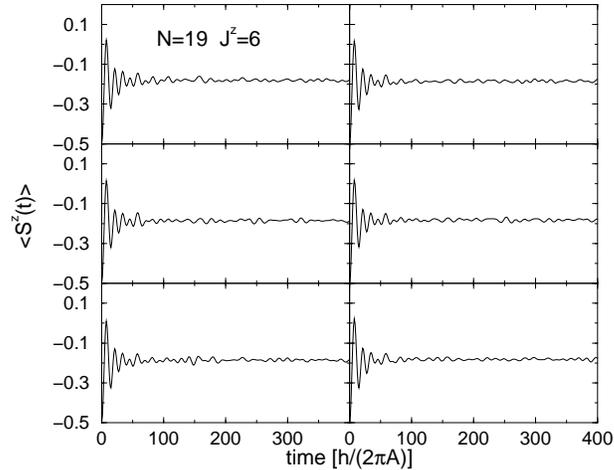}} 
\caption{Electron spin dynamics for an initially randomly correlated
nuclear spin system for $N=19$ nuclear spins with a moderate degree
of polarization ($J^{z}=6$). The data of six completely independent
random realizations of the initial nuclear spin state are shown.
The resulting electron spin dynamics is practically independent
of the realization of the initial state for this type of initial condition. 
\label{dynamics2}}
\end{figure}

\begin{figure}
\centerline{\includegraphics[width=8cm]{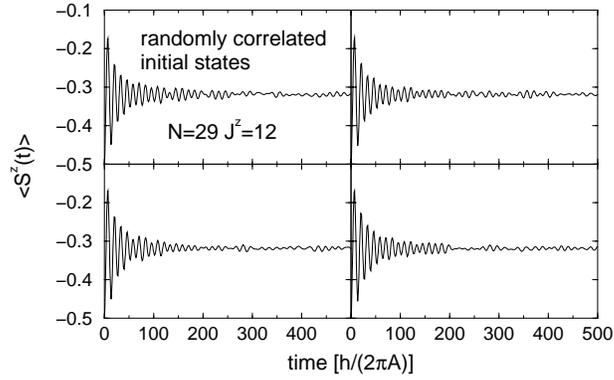}} 
\caption{Data of the same type as in Fig.~\protect{\ref{dynamics2}}
for $N=29$ nuclei and $J^{z}=12$. Again, the electron spin dynamics is
practically independent of the realization of the random initial nuclear state.
\label{dynamics3}}
\end{figure}

When the nuclear spin system is initially in a randomly correlated
state the time evolution of $\langle S^{z}(t)\rangle$ is very
reproducible in the sense that it depends only very weakly on the
particular representation of the initial random state. This is
illustrated in Figs.~\ref{dynamics2} and \ref{dynamics3}
where the results of different randomly correlated initial
nuclear spin states are compared for two different
system sizes and degrees of polarization.

\begin{figure}
\centerline{\includegraphics[width=8cm]{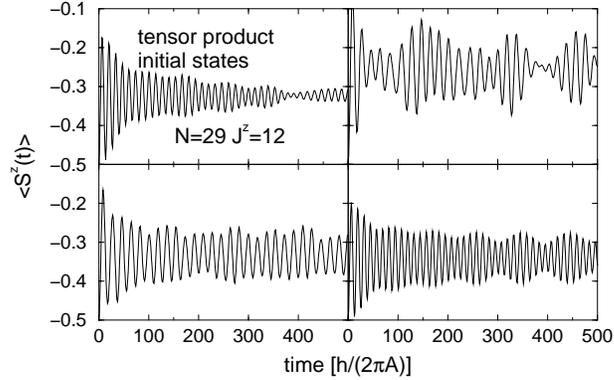}} 
\caption{Electron spin dynamics for $N=29$ nuclei and $J^{z}=12$.
Here the initial state of the nuclear spins is given by individual
tensor product states. Different initial tensor product states
clearly lead to a significantly different time evolution of the electron spin.
As seen in Fig.~\protect{\ref{dynamics3}} this is strikingly different from
randomly correlated initial conditions.
\label{dynamics4}}
\end{figure}

This behavior of randomly correlated initial states sharply
contrasts with the time evolution of simple tensor product nuclear
spin state. Fig.~\ref{dynamics4} shows the time
evolution of the electron spin for two initial tensor product
states for the same system size and degree of polarization as
in Fig.~\ref{dynamics3}. 
The comparison of these two figures
demonstrates the significant difference in the electron spin dynamics
for these two types of initial conditions in the nuclear system.
In the case of tensor product initial states the time evolution depends
significantly on the concrete initial condition, and
the decay of the electron spin occurs typically more
slowly than in the case of an initially randomly correlated
nuclear spin system. 

\subsubsection{Spin decay and quantum parallelism}
\label{quantpar}

\begin{figure}
\centerline{\includegraphics[width=8cm]{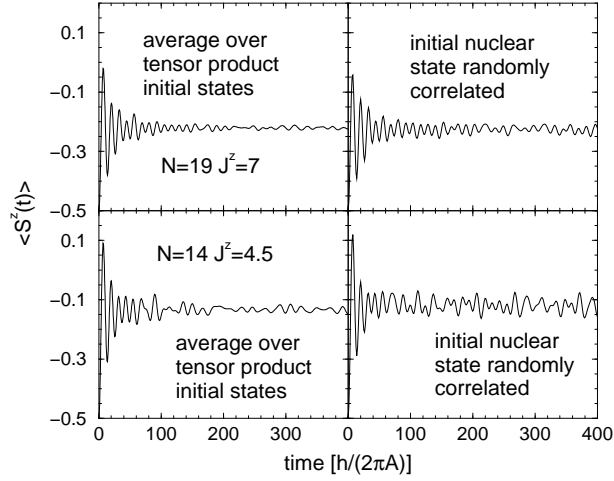}} 
\caption{The left two panels show the electron spin dynamics 
{\em averaged over all possible initial tensor product states}
for two different system sizes and degrees of polarization.
The right two panels show the corresponding data
for a single randomly correlated initial condition for the nuclear system.
As the comparison shows, the time evolution for the randomly correlated
nuclear spin system closely mimics the average over all tensor
product initial states. 
\label{dynamics5}}
\end{figure}

In the two left panels of Fig.~\ref{dynamics5} we show the time
evolution of the electron spin averaged over all nuclear tensor
product states for two different system sizes and
initial polarizations of the nuclear spins. The two right panels
show the corresponding data for a single randomly correlated initial
nuclear state.
Comparing those plots one
sees that this data is very close to the time evolution of a
randomly correlated state. This observation is also made for other
system sizes and degrees of polarization and constitutes an
example of {\em quantum parallelism} \cite{Nielsen00}: 
The time evolution of each
initially uncorrelated (and therefore classical-like) nuclear
state is present in the evolution of a linear superposition of all
such states. In other words, the time evolutions of all
uncorrelated classical-like states are performed in parallel
in the time evolution of the randomly correlated state.
An experimental consequence of this observation is
that if the electron spin dynamics would be detected on an array
of independent quantum dots one could not distinguish whether the
nuclear spin system in each dot was initially randomly correlated
or in an uncorrelated tensor product state. In other words, the
spin dynamics of a  randomly correlated pure state of the nuclear
system in a single dot cannot be distinguished from a mixed state
of an ensemble of dots.

\begin{figure}
\centerline{\includegraphics[width=8cm]{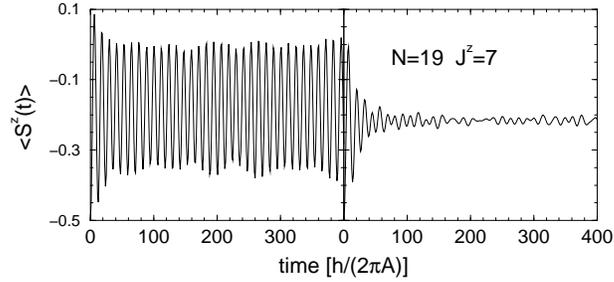}} 
\caption{Time evolution of $\langle S^{z}(t)\rangle$ for two types of initially
randomly correlated nuclear spin states. In the left panel the
amplitudes $\alpha_{T}$ are restricted to have non-negative real
and imaginary part, while in the right panel they have all the
same modulus but completely random phases.
\label{dynamics6}}
\end{figure}

As seen above,
for randomly correlated initial nuclear states the time evolution
of the electron spin does practically not depend on the concrete realization
of the random nuclear state and mimics closely the
average over all tensor product initial conditions.
This observation relies on the cancellation of off-diagonal
terms $\alpha_{T}^{\ast}\alpha_{T'}\langle \Downarrow,T|\vec
S(t)|\Downarrow,T'\rangle$, $T\neq T'$, due to the randomness in
the phases of the coefficients $\alpha_{T}$. In this sense our
system has a self-averaging property. This can be checked
explicitly by reducing this randomness. The left panel of
Fig.~\ref{dynamics6} shows the time evolution of a randomly correlated
state where the amplitudes $\alpha_{T}$ are restricted to have a
non-negative real and imaginary part. This time evolution turns
out to be similarly reproducible as before, i.e. it does not
depend on the concrete realization of the initial random state,
but it is clearly different from the former case since the
cancellation of off-diagonal contributions is inhibited.
For comparison we show in the right panel of Fig.~\ref{dynamics6}
data where
the amplitudes in the initial nuclear spin state have a random
phase but are restricted to have the same modulus. Here the proper
averaging process takes place again.

The results described so far were
obtained in certain subspaces of $J^{z}$ and for the form of
coupling constants $A_{i}$ as induced by the quantum dot geometry.
However, our findings do not depend on these choices. We have also
performed simulations were the initial state has overlap in the
full Hilbert space. For a randomly correlated initial nuclear spin
state the only difference is that now also transverse components
$\langle S^{x}(t)\rangle$, $\langle S^{y}(t)\rangle$ of the
electron spin evolve. However, these are tiny in magnitude and
oscillate around zero. For an initial tensor product states these
transverse components can become sizable, and the time evolution
again strongly depends on the concrete initial tensor product
state. Moreover, as mentioned earlier, the exact form of the
coupling constants is also not crucial as long as they are
sufficiently inhomogeneous. For instance, we obtain qualitatively
the same results if we choose the coupling parameters randomly
from a uniform distribution.

\subsection{Decoherence and the generation of entanglement}
\label{entanglement}

In circumstances of quantum information processing the decay of a
qubit is usually viewed as some 'decoherence' process due to the
environment attacking the quantum information. As seen above, the
spin decay is generically slower if the spin environment is
initially in a uncorrelated state. This finding suggests that it
is advantageous for protecting quantum information to disentangle
the environment that unavoidably interacts with the qubit system.

A 'decoherence' process of the above kind can be viewed as the
generation of entanglement between a qubit and its environment.
The system investigated here provides an illustrative example for
this statement. The entanglement in the total state
$|\psi(t)\rangle$ between the central electron spin and its
environment can be measured by the von-Neumann entropy of the
partial density matrix where either the electron or the
environment has been traced out from the pure-state density matrix
$|\psi(t)\rangle\langle\psi(t)|$ \cite{Bennett96}. Tracing out the
nuclear system we have
\begin{equation}
\rho_{el}(t)=\left(
\begin{array}{cc}
\frac{1}{2}+\langle S^{z}(t)\rangle & \langle S^{+}(t)\rangle\\
\langle S^{-}(t)\rangle & \frac{1}{2}-\langle S^{z}(t)\rangle
\end{array}
\right) \, .
\end{equation}
This matrix has eigenvalues $\lambda_{\pm}=1/2\pm|\langle\vec
S(t)\rangle|$, and the measure of entanglement reads
$E(|\psi(t)\rangle)=
-\lambda_{+}\log\lambda_{+}-\lambda_{-}\log\lambda_{-}$. Thus, the
formation of expectation values $|\langle\vec S(t)\rangle|\neq
1/2$ (or, in the case of fixed $J^{z}$, just $|\langle
S^{z}(t)\rangle|\neq 1/2$), is a manifestation of the entanglement
between the electron spin and the nuclear spin system. The maximum
entanglement, $E=\log 2$, is achieved if the electron spin has
decayed completely as measured by the expectation values of its
components, $\langle\vec S(t)\rangle=0$. The generation of quantum
entanglement between the electron spin and the nuclear spin system
signaled by a reduced value of $\langle\vec S(t)\rangle$ is a
main and crucial difference between the quantum system studied
here and its classical 'counterpart' described by a system of
Landau-Lifshitz equations. These equations can be obtained
from the Heisenberg equations of motion for the quantum system,
$\partial\vec S/\partial t=i[{\cal H},\vec S]/\hbar$,
$\partial\vec I_{i}/\partial t=i[{\cal H},\vec I_{i}]/\hbar$,
by performing expectation values of both hand sides
within spin-coherent states and assuming that the expectation values
of all operator products factorizes to products of expectation values.
This procedure becomes exact in the classical limit
\cite{Schliemann98}. The resulting equations do not contain 
operators any more but just describe the dynamics of three-component
vectors (classical spins) of fixed length.  We have performed simulations 
of such a classical spin system
by solving the Landau-Lifshitz equation via the fourth-order Runge-Kutta
scheme. As a result, the central classical spin performs an irregular
chaotic motion which does not show any similarity to the results
for the quantum spin-$\frac{1}{2}$ case.
In particular all qualitative features of 
quantum effects such as the generation of entanglement (signaled by
a decay of spins as measured by their expectation values) are not
present in such a time evolution.
Therefore the Landau-Lifshitz equation 
provides only a rather poor description of the
underlying quantum system.

We finally consider the nuclear spin correlator $C(t)=\langle
I^{z}(t)I^{z}(0)\rangle$, $\vec I=\sum_{i}\vec I_{i}$, which can
be measured directly by local NMR like measurements such as
magnetic resonance force microscopy \cite{Suter02}. In a subspace
of given $J^{z}$ and the electron spin pointing downwards
initially this quantity reads $C(t)=(J^{z}-\langle
S^{z}(t)\rangle)(J^{z}+1/2)$. A realistic initial state will have
its dominant weight in a series of subspaces with neighboring
$J^{z}$ centered around some value. Then the time evolution of
$\langle S^{z}(t)\rangle$ is very similar in these subspaces, and
the dynamics of the total nuclear spin can be mapped out by
measuring the electron spin and vice versa.

\subsection{Averaging over nuclear configurations. 
Dephasing time for an ensemble of dots} 
\label{ensemble}

In section \ref{elementary} we have seen 
(see also section \ref{pert}) that the  decay
of  $\langle S^z(t)\rangle$ for the initial tensor product state occurs 
starting from the time 
 $ t > \hbar N/A $,  
with $\hbar N/A \simeq 10^{-6} s$ in GaAs dots. On the other hand,
the electron spin precesses in the net nuclear field 
with the characteristic period 
$\omega_N^{-1}\simeq \hbar\sqrt{N}/A\simeq 10^{-8}\div 10^{-9} {\rm s}$.
Thus we see that the electron spin undergoes many  precessions in a given
nuclear field $\vec h_{N}=\sum_{i}A_{i}\langle\vec I_{i}\rangle$
(for a given nuclear configuration)
before decoherence sets in due to the non-uniform 
hyperfine couplings $A_{i}$.  
This  behavior changes dramatically when we average over  nuclear
configurations \cite{Khaetskii02a}.
Let us average $C_{T}(t)=\langle T|S^{z}(t)-S^z(0)|T\rangle$ over all initial 
tensor product nuclear configurations $|T\rangle$. 
For that purpose we  calculate  $C_{T}(t)$
exactly by treating the nuclear field  purely classically, i.e. as a 
c-number.  Then we obtain,  
\begin{equation}
C_{T}(t)=  \frac{h_{N\perp}^2}{2 h_N^2} (1- \cos(h_{N}t)), 
\label{classic}
\end{equation} 
where $h_N = \sqrt{h_{Nz}^2 + h_{N\perp}^2}$ is the nuclear field, with 
$ \/ h_{N\perp}^2= h_{Nx}^2 + h_{Ny}^2$. Again, the value of  
$h_N$  corresponds to a given nuclear tensor product state
$|T\rangle$.  We average 
Eq.(\ref{classic}) over a Gaussian distribution for $h_N$, i.e. over 
$P(h_N) \propto
\exp(- 3h_N^2/2\omega_N^2)$.
Defining $C_{cl}(t)=\int dh_N P(h_N) C_T(t)$, we obtain 
\begin{equation}
C_{cl}(t)= \frac{1}{3} [1+ ( \frac{\omega_N^2
t^2}{3} -1)\, e^{-\omega_N^2 t^2/6}].
\label{exact}
\end{equation} 
Thus we get a very rapid (Gaussian) decay  of $C_{cl}(t)$ for 
$t\gg \omega_N^{-1}$ which means that the 
dephasing time is
$\sqrt{N}/A$. From the above equation we see 
that $\langle S^z(t)\rangle$ saturates at 1/3 of its
initial value of $-1/2$. An important approximation used here is the classical
treatment of the nuclear field as a c-number (not an operator). 
The important qualitative point
we want to illustrate here is the difference between the decoherence of 
an electron in an individual quantum dot compared to the case of
an emsemble of dots. 

Investigations similar in spirit to the above
considerations were performed by Merkulov, Efros, and Rosen \cite{Merkulov02} 
who have also considered the problem 
of the electron spin dephasing due to the hyperfine interaction for an 
ensemble of dots. As was already mentioned above, the field exerted on the 
electron spin by hyperfine interaction with 
surrounding nuclei is typically much larger at $N\gg 1$ than the field the 
electron provides to an individual nucleus. Following this observation
 Merkulov {\it et al.} \cite{Merkulov02}  observed several time scales  with 
different decay laws. 
At times shorter than $\hbar N/A$ they have considered an approximation
where the nuclear spins are assumed to be static on the typical time scale
of the electron spin dynamics. That is, in an individual quantum dot the
electron spin dynamics is approximated by a coherent rotation in the
hyperfine field provided by the ``frozen'' nuclear spin configuration.
Dephasing of the electron spin is then obtained by averaging over an ensemble
of dots with individual nuclear spin configurations. 
Then they obtained essentially the same formula, see Eq.(\ref{exact}), 
with the same time scale- $\hbar\sqrt{N}/A$.    
The resulting dephasing times $T_{2}^{*}$ obtained in \cite{Merkulov02}
are in reasonable agreement with experiments \cite{Gupta99,Epstein01}.

Note that at later time scale ($t \gg \hbar N/A$) 
the authors of \cite{Merkulov02} 
obtained further slow electron spin decay by considering the  variations of 
the nuclear field direction when the magnitude of the field is conserved, 
which is correct at $N\gg 1$.  

\subsection{Further approximate studies of electron spin dynamics}
\label{approx}

The exact solution for the electron spin dynamics under the hyperfine 
coupling (\ref{defham}) obtained in Ref.~\cite{Khaetskii02a} and 
described briefly in section \ref{elementary} is unfortunately restricted
to the case of an initially fully polarized nuclear spin system. The
numerical approach described in section \ref{numdyn} allows for a
(quasi-)approximation-free treatment of systems with an arbitrary degree
of initial nuclear polarization, but is, in particular at lower
polarization, restricted to rather small systems. Therefore, in order to
investigate the physically most interesting case of larger systems with
initially moderate or even very low nuclear polarization, one needs to 
resort to approximations. In this section we review several
recent approaches. 

\subsubsection{Perturbation theory}
\label{pert}

The contribution to the hyperfine Hamiltonian (\ref{defham}) 
coupling to the $z$-component of the electron spin, 
\begin{equation}
{\cal H}_{0}=g\mu_{B}S^{z}B+S^{z}\sum_{i}A_{i}I_{i}^{z}
\label{hamzero}
\end{equation}
is diagonal in a basis of tensor product states with respect to the $z$-axis
as formulated in Eq.~(\ref{updowntensor}). The remaining part of the
Hamiltonian,
\begin{equation}
V=\sum_{i}\frac{A_{i}}{2}\left(S^{+}I_{i}^{-}+S^{-}I_{i}^{+}\right)\,,
\label{hampert}
\end{equation}
was treated in Refs.~\cite{Khaetskii02a,Khaetskii02b} as a perturbation
to ${\cal H}_{0}$. Again we shall concentrate on the case of nuclear spins 
of length $1/2$. Assuming the system to be initially ($t=0$) in a tensor 
product state with respect to the $z$-axis (cf.~(\ref{updowntensor})) with the 
electron spin as before pointing downwards, the lowest nonvanishing
contribution to the electron spin dynamics in time-dependent
perturbation theory is of second order in $V$.
Specifically, one finds
\begin{equation}
\langle S^{z}(t)\rangle=-\frac{1}{2}
+2\sum_k \frac{|V_{ik}|^2}{\omega_{ik}^2} 
\left(1-\cos(\omega_{ik}t)\right)\,,
\label{pertlo}
\end{equation}
where the summation goes over all intermediate  states $|k\rangle$ which are tensor product states of the form (\ref{updowntensor}).
$V_{ik}$ is the matrix
element of $V$ between an intermediate state $|k= \Uparrow, 
\{...., I^k_z = -1/2,... \} \rangle$ and the initial state $|i=\Downarrow, \{..., I^k_z = +1/2,... \}\rangle$, and
$\omega_{ik}=(\varepsilon_{i}-\varepsilon_{k})/\hbar$, where 
$\varepsilon_{i}$, $\varepsilon_{k}$ are the 
eigenvalues of $|i\rangle$ and $|k\rangle$,
respectively, with respect to ${\cal H}_{0}$. 

Evaluating this lowest-order contribution for a large ($N\gg 1$) unpolarized 
system, one finds an universal power law for the decay
of $\langle S^{z}(t)\rangle$: For times large compared to $\hbar N/A$
$|\langle S^{z}(t)\rangle|$ decays as $t^{-3/2}$. This is a central finding
within this perturbative approach and agrees with the perturbative limit 
(i.e. large Zeeman field) of the 
exact solution in the fully polarized case discussed in 
section \ref{elementary}.  Note that for a weak Zeeman field  $\epsilon_z 
< A/\sqrt{N} $ the part of the 
 electron spin state which decays is of the order of the initial value, 
in contrast to the fully polarized case where this part is of order $1/N$, 
see Fig.~\ref{exsol1}. 

The perturbative approach has the following shortcomings: Clearly,
${\cal H}_{0}$ and $V$ are of the same order of magnitude if one does
not apply a very large external magnetic field. Therefore, there is in general
no small parameter justifying a perturbative expansion concentrating
on low orders, and one would need to sum over higher orders, provided
that this perturbative series has sufficient convergence properties.
In higher orders, however, one encounters increasing divergences due to
vanishing denominators in the perturbative contributions \cite{Khaetskii02a}. 
Therefore,
the findings obtained from the low-order result (\ref{pertlo}) might
appear not very reliable. Indeed, from the exact solution in the fully
polarized case \cite{Khaetskii02a,Khaetskii02b} (cf. section \ref{elementary})
one finds a different (not a power) law for spin decay in the limit of 
low external magnetic field. 
However, there is a reasonable hope that the basic conclusion from the 
low-order
perturbative approach is still correct, namely that a non-uniform 
hyperfine coupling leads to a non-exponential spin decay. 

Moreover, the perturbative approach can, for technical reasons, only
deal with initial states where the nuclear spin system is in a tensor
product state. As seen in section \ref{numdyn} the behavior of such 
type of initial states depends significantly on the particular initial 
condition, in contrast to the behavior of randomly correlated initial states.

\subsubsection{Studies using Markovian approximations to the nuclear spin 
dynamics}
\label{Markovian}

Saykin, Mozyrsky, and Privman \cite{Saykin02}, and de Sousa and Das Sarma
\cite{deSousa02a,deSousa02b} have performed investigations using, among
other simplifying assumptions, Markovian approximations to describe the 
dynamics of the nuclear spins. 

In Ref.~\cite{Saykin02} the situation of an electron bound to a $^{31}{\rm P}$
donor in a silicon matrix was studied. The electron spin interacts
with the central $^{31}{\rm P}$ and the surrounding $^{29}{\rm Si}$
nuclear spins. Then the authors formulate a master equation for the
reduced electron density matrix where the nuclear spin dynamics is governed by
a Markov process. Moreover, the total density matrix is assumed to be
separable at all times with respect to the subsystems 
given by the electron spin and the nuclear spin bath, 
and the density operator of the 
latter system is assumed to be time-independent. 
Due to the first assumption, decoherence due to
the formation of entanglement as described in section \ref{entanglement}
is excluded since the total density matrix is taken to be always 
separable \cite{Nielsen00}. 
As a result, the authors find an exponential decay of the elements
of the reduced density matrix to their equilibrium values, where the 
longitudinal relaxation time $T_{1}$ and the transversal
dephasing time $T_{2}$ (with respect to the direction of a weak external field)
fulfill the relation $T_{1}=T_{2}/2$.  
However, as already explained in section \ref{elementary}, the application of
the Markov approximation in this situation is not justified, 
which might explain that this result differs qualitatively
from the findings from the exact solution in the case of an initially
fully polarized nuclear spin system (cf. section \ref{elementary})
and the perturbative result for the unpolarized case (cf. section \ref{pert}),
where a power law decay was found.

In Refs.~\cite{deSousa02a,deSousa02b} the Hamiltonian (\ref{defham}) was
studied under the assumption of strong external magnetic field
coupled to the electron spin and suppressing the spin-flip terms.
Therefore, the hyperfine interaction is approximated by (\ref{hamzero}).
This approach is applicable at an external Zeeman magnetic field which is 
much larger than the internal nuclear field, i.e. 
$\varepsilon_z \gg \hbar\omega_N \simeq A/\sqrt{N}$. We recall that the part 
of the initial electron spin
state which decays due to inhomogeneous hyperfine coupling is small under the 
above conditions, this part is of the order of 
$(\hbar\omega_N/\varepsilon_z)^2 \ll 1 $, see 
Refs.\cite{Khaetskii02a,Khaetskii02b}. 
The authors of Refs.~\cite{deSousa02a,deSousa02b} then introduced 
the nontrivial dynamics through the dipolar interaction
among the nuclear spins. This interaction is taken into account
in a truncated version neglecting terms that change the total spin
component in the direction of the external field, consistent with the
earlier approximation. Thus, in this approximation 
the dipolar dynamics among the nuclear spins
induce the decay of the electron spin. As a further assumption, the 
nuclear spin system is then approximated as a fluctuating field
generated by a Markov process.
After performing a detailed mathematical analysis of this resulting
effective model \cite{deSousa02b} the authors conclude that the electron
spin decay will occur on essentially the same time scale as the
dipolar interaction among the nuclear spins, i.e. 
$T_{n2}\simeq 10^{-4}\div 10^{-5}s$.
Given the various assumptions leading to this effective model, this is a 
very natural result.

Applying their findings to the situation of an electron bound to a
$^{31}{\rm P}$ donor embedded in an Si crystal, the authors find
an interesting dependence of the spin memory time $T_{M}$ (as measured
by spin-echo experiments) on the direction of the external field
relatively to the crystal. This effect is induced by the
directional dependence of the dipolar interaction and provides
a possibility to experimentally probe these results and the
underlying assumptions \cite{Tyryshkin03}.

\subsection{Further studies and developments}
\label{further}

We now summarize further, mostly theoretical, work relevant to the issue
of hyperfine interaction between an electron spin bound to a quantum dot
and surrounding nuclear spins.

The relaxation rate of the longitudinal electron spin component, i.e. 
$1/T_1$, which is due to  the 
interplay between hyperfine interaction and dissipative phonon processes
was studied by Erlingsson, Nazarov, and Falko \cite{Erlingsson01}, 
and by Erlingsson and Nazarov \cite{Erlingsson02}. 
The approach in Ref.~\cite{Erlingsson02}
is a semiclassical one introducing an internal field due to the nuclear
spins which acts on the electron spin in addition to an external magnetic 
field. The observed relaxation rate is very small, though this mechanism can 
prevail over the mechanism which is the interplay of the spin-orbit 
interaction and spin-independent interaction with phonons \cite{Khaetskii01}, 
it happens at very low external magnetic field when the corresponding 
$T_1$ time is order of $100s$. 
The analogous problem of the calculation of the $T_2$ time which is  due to 
the combined effect of hyperfine interaction
and phonon processes was  considered recently by 
Semenov and Kim \cite{Semenov03b}. 
These authors noticed that due to thermal fluctuations the electron can 
make spin-conserving transitions between different orbital states. 
Then because of either the energy dependence of the g-factor or 
different nuclear fields seen by electron spin in different orbital 
states these fluctuations lead to electron spin decoherence 
(since the precession frequencies are different in different states).  
These mechanisms are only important at relatively high temperatures since 
they are exponentially suppressed at temperatures much smaller than the 
energy distance between  the neighbouring  
orbital states. For example, in the case of combined effect of 
hyperfine interaction
and phonon processes for typical GaAs quantum dots the decoherence time is of 
order of $10s$ at temperature 1K. 

Lyanda-Geller, Aleiner, and Altshuler \cite{Lyanda-Geller02} have
investigated the dependence of the relaxation rate of nuclear spins in a 
quantum dot should on the electronic state of the dot and concluded that 
effects of Coulomb blockade should also be observable in nuclear relaxation 
in such systems.

A proposal for using the nuclear spin system as a long-lived
memory for information originally contained in the electron spin qubit
was put forward by Taylor, Marcus, and Lukin \cite{Taylor03}.
In a related study, Imamoglu, Knill, Tian, and Zoller \cite{Imamoglu03}
have proposed an all-optical scheme for polarizing the nuclear 
spins by manipulating the electron spin. 

Saikin and Fedichkin \cite{Saikin02}
have investigated the influence of hyperfine interactions
on gate operations within the Si:P quantum computing proposal due to
Kane \cite{Kane98}. 

The possibility of nuclear spins forming an effective quantum dot
confining electrons through hyperfine interaction was proposed recently by 
Pershin \cite{Pershin03b}.

Theoretical studies on the decoherence of a two level systems coupled 
to surrounding spins were also presented recently by Frasca \cite{Frasca02}.
These studies stem from a somewhat different context but are similar in
spirit the  ones reported on in this article.
We also mention a recent numerical study by Dobrovitski 
{\sl et al.} \cite{Dobrovitski01} on spin dynamics stressing the role of
entropy. There a central spin is coupled inhomogeneously to an
essentially non-interacting spin environment where an Ising-like coupling 
was used. To allow for nontrivial dynamics the authors 
introduced a magnetic field perpendicular to the $z$-direction of
the Ising coupling. In a very recent paper \cite{Dobrovitski03} the same 
authors have numerically studied
the damping of quantum oscillations in the system of two central spins.  
These central spins are coupled by strong Heisenberg exchange, this system  
in turn is coupled to a spin bath (which can be a nuclear system) through 
inhomogeneous Heisenberg interaction, a scenario smilar to the one 
studied in \cite{Loss98}. The state of the bath is initially a 
random superposition of all possible basis states. Assuming the coupling 
constants $A_k$ are random, in the system of $N=13$ bath spins  the authors 
observed a two-step decoherence process. Initially decoherence is very fast 
and after the first step the oscillations of $z$-components of the central 
spins $S^z_1(t)$ and $S^z_2(t)$ persist and decay very slowly. 
It is interesting that the authors have managed to fit the first step fast 
decay by Eq.(\ref{exact}), see Sec.\ref{ensemble}, which probably 
confirms the self-averaging occurring for randomly correlated  initial 
nuclear states described in Sec.\ref{quantpar}. The second slow step of the 
decoherence process is presumably the analogue of the decay described in 
sections \ref{elementary} and \ref{pert}.   
   
\section{Conclusions and outlook}
\label{concl}

We have reviewed the recent literature  
on electron spin dynamics in semiconductor nanostructures due to
hyperfine interaction with surrounding nuclear spins. 
This issue is of particular interest with respect to several
proposals for quantum information processing in solid state systems.
Although the basic Hamiltonian (\ref{defham}) looks rather simple it
describes an intricate many-body problem which in general does not seem
to allow for an analytical solution.

In the case of an initially fully polarized nuclear spin system an
exact analytical solution for the spin dynamics can be found. For
not completely polarized nuclei approximation-free results can only be 
obtained numerically in small model systems. We have compared these
exact results with findings from several approximation strategies
such as perturbation theory and (semi-)classical approximations to the
nuclear spin dynamics including Markovian approximations.

Many of those approximation are not particularly well controlled, and they
suppress important features of the full quantum system. For instance,
the pronounced dependence of the dynamics of the electron spin on
the type of initial condition for the nuclear system is not reproduced
by any of the approximation strategies which have appeared so far.
Therefore, the most obvious direction for future work is the
development of possibly systematically
controlled approximation techniques which reproduce important
features of the full quantum dynamics and allow reliable predictions
for realistic systems. A possible but presumably techically quite 
complicated route for such future work is the Bethe ansatz solution
outlined in section \ref{integ}. 
Such progress toward more reliable predictions
for experimentally relevant situations is especially desirable because of
the importance of this issue to several proposals for quantum information
processing in a semiconductor environment, see in particular 
Refs.~\cite{Loss98,Kane98}, and
possibly also for other scenarios in the emerging field of spin
electronics \cite{Wolf01,Awschalom02}.

We thank B.~L. Altshuler, W.~A. Coish, S.~I. Erlingsson, Al.~L. Efros
L. Glazman, F. G\"ohmann, V.~N. Golovach, Y.~N. Nazarov, 
and D. Saraga for collaborations
or/and discussions. This work was supported by NCCR Nanoscience, the Swiss NSF,
DARPA, and ARO.

\newpage

\begin{table}
\begin{center}
\begin{tabular}{|c|c|c|c|}
\hline
  &  ${\rm natural}\atop{{\rm abundance}[\%]}$& $I$ & $\mu_{I}$ \\ 
\hline
$^{9}{\rm Be}$ & 100 & 3/2 & --1.1776 \\
\hline
$^{10}{\rm B}$ & 19.78 & 3 & +1.8007 \\
$^{11}{\rm B}$ & 80.22 & 3/2 & +2.6885 \\
\hline
$^{13}{\rm C}$ & 1.11 & 1/2 & +0.7024 \\
\hline
$^{14}{\rm N}$ & 99.63 & 1 & +0.4036 \\
$^{15}{\rm N}$ & 0.37 & 1/2 & --0.2831 \\
\hline
$^{27}{\rm Al}$ & 100 & 5/2 & +3.6414 \\
\hline
$^{29}{\rm Si}$ & 4.70 & 1/2 & --0.5553 \\
\hline
$^{31}{\rm P}$ & 100 & 1/2 & +1.1317 \\
\hline
$^{33}{\rm S}$ & 0.76 & 3/2 & +0.6433 \\
\hline
$^{67}{\rm Zn}$ & 4.11 & 5/2 & +0.8754 \\
\hline
$^{69}{\rm Ga}$ & 60.4 & 3/2 & +2.016 \\
$^{71}{\rm Ga}$ & 39,6 & 3/2 & +2.562 \\
\hline
$^{73}{\rm Ge}$ & 7.76 & 9/2 & --0.8792 \\
\hline
$^{75}{\rm As}$ & 100 & 3/2 & +1.439 \\
\hline
$^{77}{\rm Se}$ & 7.58 & 1/2 & +0.534 \\
\hline
$^{111}{\rm Cd}$ & 12.75 & 1/2 & --0.5943 \\
$^{113}{\rm Cd}$ & 12.26 & 1/2 & --0.6217 \\
\hline
$^{113}{\rm In}$ & 4.28 & 9/2 & +5.523 \\
$^{115}{\rm In}$ & 95.72 & 9/2 & +5.534 \\
\hline
$^{115}{\rm Sn}$ & 0.35 & 1/2 & --0.918 \\
$^{117}{\rm Sn}$ & 7.16 & 1/2 & --1.000 \\
$^{119}{\rm Sn}$ & 8.58 & 1/2 & --1.046 \\
\hline
$^{121}{\rm Sb}$ & 57.25 & 5/2 & +3.359 \\
$^{123}{\rm Sb}$ & 42.75 & 7/2 & +2.547 \\
\hline
$^{123}{\rm Te}$ & 0.87 & 1/2 & --0.7357 \\
$^{125}{\rm Te}$ & 6.99 & 1/2 & --0.8871 \\
\hline
$^{199}{\rm Hg}$ & 16.84 & 1/2 & +0.5027 \\
$^{201}{\rm Hg}$ & 13.22 & 3/2 & --0.5567 \\
\hline
$^{207}{\rm Pb}$ & 22.6 & 1/2 & +0.5895 \\
\hline
$^{209}{\rm Bi}$ & 100 & 9/2 & +4.080 \\
\hline
\end{tabular}
\caption{Natural abundances, total nuclear spin quantum numbers $I$, and
nuclear magnetic moments $\mu_{I}$ of spin-carrying stable 
nuclei relevant to semiconductor
materials. The values for $\mu_{I}$ are given in units of the nuclear
magneton $\mu_{N}=e\hbar/2m_{p}$ where $m_{p}$ is the proton mass.
The data is adopted from the 
{\em American Institute of Physics Handbook}, Third Edition, McGraw-Hill, 1972.
\label{table1}}
\end{center}
\end{table}

\end{document}